\documentclass[aps,prd,preprintnumbers,showpacs,nofootinbib,onecolumn]{revtex4}
\usepackage{graphicx}
\usepackage{bm}
\usepackage{amsmath}
\usepackage{amssymb}
\usepackage{amsthm}
\usepackage{color}
\usepackage[usenames,dvipsnames]{xcolor}
\usepackage{latexsym}
\usepackage{epsfig}
\newcommand{\lp}{\left(}
\newcommand{\rp}{\right)}
\newcommand{\lb}{\left[}
\newcommand{\rb}{\right]}

\newcommand{\D}{{\rm D}}
\newcommand{\mD}{\mathcal{D}}
\newcommand{\mR}{\mathcal{R}}
\newcommand{\mQ}{\mathcal{Q}}
\newcommand{\mP}{\mathcal{P}}
\newcommand{\divA}{\nabla \cdot A}
\newcommand{\hX}{\hat{X}}
\newcommand{\hGam}{\hat{ \Gamma}}
\newcommand{\hnabla}{\hat{ \nabla}}
\newcommand{\hR}{\hat{ R}}

\newcommand{\hG}{\hat{ \mathcal G}}

\newcommand{\Lag}{{\mathcal L}}
\newcommand{\Od}{{\cal O}}
\newcommand{\be}{\begin{equation}}
\newcommand{\ee}{\end{equation}}
\newcommand{\lsim}   {\mathrel{\mathop{\kern 0pt \rlap
  {\raise.2ex\hbox{$<$}}}
  \lower.9ex\hbox{\kern-.190em $\sim$}}}
\newcommand{\gsim}   {\mathrel{\mathop{\kern 0pt \rlap
  {\raise.2ex\hbox{$>$}}}
  \lower.9ex\hbox{\kern-.190em $\sim$}}}
\newcommand{\bw}{\begin{widetext}\begin{equation}}
\newcommand{\ew}{\end{equation}\end{widetext}}
\newcommand{\beq}{\begin{equation}}
\newcommand{\eeq}{\end{equation}}
\newcommand{\bea}{\begin{eqnarray}}
\newcommand{\eea}{\end{eqnarray}}

\newcommand{\diff}{{\rm d}}

\begin{document}

\title{Extended Gauss-Bonnet gravities in Weyl geometry}
\author{Jose Beltr\'an Jim\'enez${^a}$ and Tomi S. Koivisto${^{b,c}}$}
\affiliation{${^a}$Centre for Cosmology, Particle Physics and Phenomenology,
Institute of Mathematics and Physics, Louvain University,
2 Chemin du Cyclotron, 1348 Louvain-la-Neuve (Belgium)}
\affiliation{${^b}$Institute of Theoretical Astrophysics, University of Oslo, P.O. Box 1029 Blindern, N-0315 Oslo, Norway}
\affiliation{${^c}$Nordita, KTH Royal Institute of Technology and Stockholm University, Roslagstullsbacken 23,
SE-10691 Stockholm, Sweden}

\date{\today}

\begin{abstract}
In this paper we consider an extended Gauss-Bonnet gravity theory in arbitrary dimensions and in a space provided with a Weyl connection, which is torsionless but non-metric-compatible, the non-metricity tensor being determined by a vector field. The considered action consists of the usual Einstein-Hilbert action plus all the terms quadratic in the curvature that reduce to the usual Gauss-Bonnet term for vanishing Weyl connection, i.e., when only the Levi-Civita part of the connection is present.  We expand the action in terms of Riemannian quantities and obtain vector-tensor theories. We find that all the free parameters only appear in the kinetic term of the vector field so that two branches are possible: one with a propagating vector field and another one where the vector field does not propagate. We focus on the propagating case. We find that in 4 dimensions, the theory is equivalent to Einstein's gravity plus a Proca field. This field is naturally decoupled from matter so that it represents a natural candidate for dark matter. Also in $d=4$, we discuss a non-trivial cubic term in the curvature that can be constructed without spoiling the second order nature of the field equations because it leads to the vector-tensor Horndeski interaction. In arbitrary dimensions, the theory becomes more involved. We show that, even though the vector field presents kinetic interactions which do not have $U(1)$ symmetry, there are no additional propagating degrees of freedom with respect to the usual massive case. Interestingly, we show that this relies on the fact that the corresponding St\"uckelberg field belongs to a specific class within the general Horndeski theories. Finally, since Weyl geometries are the natural ground to build scale invariant theories, we apply the usual Weyl-gauging in order to make the Horndeski action locally scale invariant and discuss on new terms that can be added.

\end{abstract}

%\keywords{Cosmology: Dark energy, perturbations.}
\preprint{NORDITA-2014-14}

\maketitle

\section{Introduction}
	Einstein's theory of General Relativity (GR) is about to turn its first century of existence and still provides the most successful and widely accepted theory for gravity. It is based on a purely geometrical interpretation of the gravitational interaction with the metric of the spacetime playing a central role. In GR, the geometrical properties are fully determined by the metric tensor $g_{\mu\nu}$, which is itself a dynamical quantity linked to the matter content via Einstein's equations. A key ingredient of Einstein's gravity is the assumption of torsionless and metric compatible spacetime connection. These two conditions completely determine the connection to be the Levi-Civita connection of the metric. 

However, the geometry of an arbitrary manifold can be more complicated than that since, in addition to the metric, we also need to specify the connection, which a priori is a completely independent object from the metric \cite{schr}. Rather than starting with minimal field content, we can 
start with minimal assumptions and thus not postulate the affine structure of the manifold to be dictated by its geometric structure. 
%than simplicity, to assume that the connection is the one compatible with the metric and it could be treated as an independent field. 
In the Palatini formalism, the scheme is indeed such, and variations of the action are taken independently with respect to the metric and the connection \cite{palatini}. It is remarkable that the Einstein-Hilbert action in the Palatini formalism is equivalent to GR, the equation of motion for the connection then imposing its metric compatibility a posteori. However, the Palatini formulation of more general actions leads to theories that are distinct from the corresponding metric theories. 
The issue of fundamental degrees of freedom in gravitational geometry has been investigated recently also adopting various novel approaches. In the so called C-theories the spacetime connection is prescribed to be generated by a conformally related metric \cite{ctheory} while the so called bimetric variational principle allows the connection to emerge from a completely independent metric \cite{bimetric}. Biconnected spacetimes \cite{biconnection} and different tetrad formulations have been also explored in interesting studies \cite{tetrad}.

The history of nonmetricity dates back to Hermann Weyl, who considered conformally invariant theories of gravity. The freedom to choose a conformal gauge was associated with a vector field that quantifies the nonmetricity of the connection. Identifying the vector with the electromagnetic gauge field potential, the set-up provided a unification of gravity and electrodynamics \cite{gauge}. It is not clear whether conformal gravity could describe consistently our universe, the issue being still investigated \cite{conformal,Bars:2013yba}. However, the nearly scale invariant spectrum of cosmological perturbations and the fact that the Higgs mass is the only term breaking scale invariance in the Standard Model of elementary particles are very suggestive indications that this could be an actual symmetry of a more fundamental theory that has been spontaneously broken. In fact, even in theories  that exhibit  an exact scale-invariance at the classical level, the corresponding renormalization scheme will generally break it at the quantum level. At this respect, scale invariant theories have been proposed as possible resolutions to the cosmological constant and/or Higgs hierarchy problems  \cite{LambdaSI} (see however \cite{Tamarit:2013vda}).

In the present work, we are not concerned about the existence of such a fundamental theory and will simply allow the spacetime to have nonmetricity described by a Weyl-type vector field, without necessarily attempting to impose conformal invariance of the theory. This approach might be thought of as the resulting low energy effective theory of a truly scale invariant theory where this symmetry has been spontaneously broken. The nonmetricity vector field can then become a physical field that appears as an additional gravitational degree of freedom associated with the non-Levi-Civita part of the connection. It turns out that such theories can reproduce various vector-tensor theories studied in the literature and also give rise to new interactions. What we thus obtain are purely geometric interpretations for such theories, where the new vector field, instead of being introduced as an ad hoc matter field, emerges from the nonmetric nature of the spacetime structure. This provides the vector field a more profound origin and, therefore, it serves as a promising guiding principle to build new theories. Vector-tensor theories have been studied actively since the original works \cite{originalVT},  especially in view of their applications to cosmology as dark energy candidates \cite{vectorDE}, fields driving an inflationary phase \cite{vectorInflation}, as curvature perturbations generator \cite{vectorcurvaton},  or as components of modified gravities theories \cite{vectorMG}.

%see for example \cite{vector,Barrow:2012ay,Jimenez:2013qsa}.   

This paper is organised as follows. First we will briefly review Weyl geometry in Section \ref{weylg}. A very natural first starting point for constructing gravity theories is to consider Einstein-Gauss-Bonnet type actions, since they reduce to GR in the Riemannian limit and are not expected to introduce ghosts or other pathologies in the more general Weyl geometry either. We compute the resulting actions in terms of GR + vector terms for such classes of theories in Section \ref{egb}. In Section \ref{fourd} we then study some simple interesting examples in the four-dimensional case, and in Section \ref{morethanfourd} look at $d>4$. There we obtain Horndeski-type scalar theories. We promote those theories into scale-invariant ones exploiting the Weyl-type gauging in Section \ref{s:weylhorndeski}. Our findings are then briefly summarised in the concluding Section \ref{conclusions}.

\section{Weyl geometry}
\label{weylg}
The essential novelty introduced by Einstein to describe gravity in geometrical terms  was the assumption that physical phenomena occur in a (pseudo) Riemannian geometry, i.e., the spacetime curvature is not trivial. This means that there is a mismatch when parallel transporting a vector along a close path or, equivalently, the direction of a vector changes under a parallel transport. A natural generalisation introduced by Weyl soon after Einstein's General Relativity was the assumption that, not only the direction, but also the length of a vector changes under a parallel displacement. In this section we will review the basics of this Weyl geometry and give the relevant geometric equations that we will need later\footnote{For an extensive compendium of useful formulae in Weyl geometry see for instance \cite{Yuan:2013cv}.}. Weyl geometry is characterized by the presence of a non-metricity tensor which is given in terms of a vector field. More precisely, the spacetime connection (which we will denote by $\hGam^\alpha_{\beta\gamma}$ and defines the covariant derivative $\hnabla$) is not metric compatible, but it is determined by the equation
\be
\hnabla_\alpha g_{\mu\nu}=-2A_\alpha g_{\mu\nu}.
\label{nonmetricity}
\ee
Unlike in Riemannian geometries where the connection is entirely determined by  the metric tensor, in Weyl geometry the connection is provided with additional degrees of freedom ascribed to the vector field $A_\mu$. A general connection can be split into three independent pieces, namely: The usual Levi-Civita piece fully determined by the metric tensor, the non-metricity part that determines the violation of metric compatibility  and, finally, the torsion-dependent part giving the antisymmetric component of the connection. Thus, in addition to the above condition, Weyl geometry assumes a symmetric connection so that the spacetime is torsionless. 
%Thus, theories within this framework are half way between pure metric theories and theories in the Palatini formalism since only an independent vector part of the connection is present. 
Under the symmetric connection condition we can fully obtain the connection by solving (\ref{nonmetricity}), whose solution is
\be
\hGam^\alpha_{\beta\gamma}=\Gamma^\alpha_{\beta\gamma}-\left(A^\alpha g_{\beta\gamma}-2A_{(\beta}\delta^\alpha_{\gamma)}\right)\,,
\label{Wconnection}
\ee
with $\Gamma^\alpha_{\beta\gamma}$ the usual Levi-Civita connection for the spacetime metric $g_{\mu\nu}$ given by the usual formula
\be
\Gamma^\alpha_{\beta\gamma}=\frac12g^{\alpha\lambda}\Big(g_{\lambda\gamma,\beta}+g_{\beta\lambda,\gamma}-g_{\beta\gamma,\lambda}\Big).
\ee
The second piece of the spacetime connection depending on the vector field is called the Weyl connection and has tensorial transformation properties, since it can be written as the difference of two connections
\be
W^\alpha_{\beta\gamma}\equiv\hGam^\alpha_{\beta\gamma}-\Gamma^\alpha_{\beta\gamma}=-\left(A^\alpha g_{\beta\gamma}-2A_{(\beta}\delta^\alpha_{\gamma)}\right).
\ee
We can expand covariant derivatives and other geometrical objects to write them  in terms of the metric-compatible connection and the vector field. This allows to transform a given action in a Weyl geometry into a theory in Riemannian geometry but with additional matter content and interactions coming from the Weyl vector field $A_\mu$. 

The Riemann curvature will be defined as:
\be
\mR_{\mu\nu\rho}{}^\alpha\equiv\partial_\nu\Gamma^\alpha_{\mu\rho}-\partial_\mu\Gamma^\alpha_{\nu\rho}+\Gamma^{\alpha}_{\nu\lambda}\Gamma^{\lambda}_{\mu\rho}-\Gamma^{\alpha}_{\mu\lambda}\Gamma^{\nu}_{\nu\rho}.
\ee
Since the spacetime geometry is provided with non-metricity, the Riemann tensor does not have the usual symmetries of Riemannian geometry, but only the antisymmetry in the two first indices remains $\mR_{\mu\nu\rho}{}^\alpha=-\mR_{\nu\mu\rho}{}^\alpha$. This leads to an ambiguity in the definition of the Ricci tensor as the trace of the Riemann tensor, since we have three independent possibilities. The first one is the usual contraction $\mR_{\mu\nu}\equiv\mR_{\mu\alpha\nu}{}^\alpha$. Notice that, even though we are considering a torsionless connection, the presence of non-metricity will induce an antisymmetric part in $\mR_{\mu\nu}$. The second possibility is the alternative independent trace $\mQ_{\mu\nu}\equiv\mR_{\mu\nu\alpha}{}^\alpha$, which is the so-called homothetic curvature and is, by definition, an antisymmetric tensor. For symmetric connections, e.g. Weyl connections, one has that $\mQ_{\mu\nu}=\mR_{[\mu\nu]}$. Finally, the third independent contraction is the co-Ricci tensor given by $\mP_\mu{}^\alpha\equiv g^{\nu\rho}\mR_{\mu\nu\rho}{}^\alpha$. Concerning the scalar curvature, there is no ambiguity since $\mQ_{\mu\nu}$ is traceless and $g^{\mu\nu}\mR_{\mu\nu}=-\mP^\mu{}_\mu$. 

After the above digression about the different possible contractions of the Riemann tensor, we can proceed to compute them for the case of a Weyl geometry and express them in terms of of the usual Riemannian objects and the Weyl vector. The Ricci tensor is given by
\begin{eqnarray}
\mR_{\mu\nu}=R_{\mu\nu}+(d-2)\Big(A_\mu A_\nu-A^2g_{\mu\nu}\Big)-g_{\mu\nu}\divA-(d-1)\nabla_\mu A_\nu+\nabla_\nu A_\mu\,,
\end{eqnarray}
where $d$ is the spacetime dimension and $R_{\mu\nu}$ is the usual Ricci tensor associated to the Levi-Civita part of the connection, i.e., it is fully determined by the spacetime metric. The remaining independent traces of the Riemann tensor relate to the Ricci tensor as follows:
\begin{eqnarray}
\mP_{\mu\nu}&=&-\mR_{\mu\nu}-2F_{\mu\nu}\\
\mQ_{\mu\nu}&=&\mR_{[\mu\nu]}=-d\,F_{\mu\nu}\,,
\end{eqnarray}
with $F_{\mu\nu}=\partial_\mu A_\nu-\partial_\nu A_\mu$ the strength tensor associated to the Weyl vector.
As commented above and we can clearly see from the previous expressions, the Ricci scalar is unambiguously defined and can also be straightforwardly computed
\be
\mR=R-(d-2)(d-1)A^2-2(d-1)\divA.
\label{expR}
\ee
%Finally, we will need the Gauss-Bonnet term, which is given by
%\be
%\hG=\hR_{\alpha\beta\gamma\delta}\hR^{\alpha\beta\gamma\delta}-4\hR_{\alpha\beta}\hR^{\alpha\beta}+\hR^2,
%\ee
%and is the next non-trivial Lovelock invariant (thus leading to second order equations of motion). The corresponding expression reads
%with $\mathcal G$ the Gauss-Bonnet term corresponding to the Levi-Civita connection $\Gamma$. 
To end this brief introduction to Weyl geometry, it is worth mentioning that the usual motivation to consider such spacetimes is the introduction of a local conformal invariance, since the non-metricity condition given in (\ref{nonmetricity}) is nothing but the definition of a covariant gauge derivative that transforms appropriately under a local conformal transformation,  being $A_\mu$ the corresponding gauge field. It is very easy to check that such a condition is invariant under $g_{\mu\nu}\rightarrow e^{2\Lambda(x)}g_{\mu\nu}$ simultaneously with $A_\mu\rightarrow A_\mu-\partial_\mu\Lambda(x)$, which means that one can define the covariant derivative 
\be
\mD_\mu g_{\alpha\beta}\equiv (\hnabla_\mu+2A_\mu)g_{\alpha\beta}\,,
\ee
which transforms covariantly under the mentioned transformation. Thus, the condition defining the Weyl geometry (\ref{nonmetricity}) is equivalent to setting $\mD_\mu g_{\alpha\beta}=0$ and is, of course, conformally  invariant. This covariant derivative allows to construct actions with local conformal invariance and has been very extensively explored in the literature. In the next section, we do not aim to construct this type of Weyl-invariant theories, but we simply assume that the spacetime has a non-metricity tensor of the Weyl form and study its consequences. However, we will come back to this issue in section \ref{s:weylhorndeski} where we will make use of this fact to construct a general theory with local Weyl invariance.

\section{Extended Gauss-Bonnet gravity in Weyl geometry}
\label{egb}
The aim of the present work is to explore gravitational theories beyond the simplest Einstein-Hilbert action within the framework of Weyl geometries. In particular, we wish to consider actions containing geometric invariants quadratic in the curvature Riemann tensor. In principle, there is a high number of different invariants that one can construct by using different contractions of the Riemann tensor. Higher curvature invariants were considered in \cite{Tanhayi:2012nn} in the context of Weyl-invariant theories, although not all the possible independent contractions were included.  Here, in order to restrict the number of terms in the resulting action, we shall restrict to those terms such that they reduce to the Gauss-Bonnet term in a Riemannian geometry, i.e., in the case when the Weyl connection vanishes. The reason for this restriction is the well-known fact that the Gauss-Bonnet term, being a Lovelock invariant, leads to second order equations of motion for the metric tensor and, thus, avoids the potential presence of Ostrogradski's instabilities.

The Gauss-Bonnet term for a Levi-Civita connection is given by
the precise combination $R_{\alpha\beta\gamma\delta}R^{\alpha\beta\gamma\delta}-4R_{\alpha\beta}R^{\alpha\beta}+R^2$. However, for a general connection, there are several combinations that lead to the Gauss-Bonnet term when imposing torsionless and metric-compatibility. For the Riemann tensor of an arbitrary connection $\mR_{\alpha\beta\gamma}{}^\delta$, there are 3 different combinations that reduce to $R_{\alpha\beta\gamma\delta}R^{\alpha\beta\gamma\delta}$ for a Levi-Civita connection, namely: $\mR_{\alpha\beta\gamma\delta}\mR^{\alpha\beta\gamma\delta}$, $\mR_{\alpha\beta\gamma\delta}\mR^{\alpha\beta\delta\gamma}$ and $\mR_{\alpha\beta\gamma\delta}\mR^{\gamma\delta\alpha\beta}$. These three combinations are trivially the same for the Levi-Civita connection because $R_{\alpha\beta\gamma\delta}=R_{\gamma\delta\alpha\beta}=-R_{\alpha\beta\delta\gamma}$ in such a case.  For the Ricci tensor contractions, we can in principle consider all the contractions of the Ricci, the co-Ricci and the homothetic curvature tensors. However, since the homothetic curvature is nothing but the antisymmetric part of the Ricci tensor for symmetric connections, it will not give rise to additional quadratic terms independent of those built with the Ricci tensor for a Weyl geometry. Thus, we only need to consider the 6 possible combinations of $\mR_{\mu\nu}$ and $\mP_{\mu\nu}$. Finally, the scalar curvature is unique so that the quadratic action that we seek for will have the general form:
\begin{align}
S_{\rm EGB}=\lambda\int\diff^dx\sqrt{-g}\hG \equiv \lambda\int&\diff^dx\sqrt{-g}\Big[c_1\mR_{\mu\nu\rho\sigma}\mR^{\mu\nu\rho\sigma}+c_2\mR_{\mu\nu\rho\sigma}\mR^{\rho\sigma\mu\nu}-c_3\mR_{\mu\nu\rho\sigma}\mR^{\mu\nu\sigma\rho}\\\nonumber
&\left.-4\Big(b_1\mR_{\mu\nu}\mR^{\mu\nu}+b_2\mR_{\mu\nu}\mR^{\nu\mu}-b_3\mR_{\mu\nu}\mP^{\mu\nu}-b_4\mR_{\mu\nu}\mP^{\nu\mu}+b_5\mP_{\mu\nu}\mP^{\mu\nu}+b_6\mP_{\mu\nu}\mP^{\nu\mu}\Big)+\mR^2\right].
\end{align}
where $c_{i}$ and $b_i$ are dimensionless constants and $\lambda$ has dimension $[\lambda]=4-d$.  Since we want to recover the Gauss-Bonnet term for the Levi-Civita part of the connection, we need to impose the following two conditions:
\be
c_1+c_2+c_3=\sum_{i=1}^6b_i=1.
\ee
The above action with the corresponding constraints on the parameters is the desired one for an arbitrary connection. However, for the case of a Weyl geometry, things are simpler because all the independent contractions appearing in that action only differ by a term proportional to $F_{\mu\nu} F^{\mu\nu}$ so that only one particular combination of the parameters is actually relevant and the action can be written as
%\footnote{See appendix for the details} 
\begin{eqnarray}
S_{\rm EGB}&=&\lambda\int\diff^dx\sqrt{-g}\left[\mathcal{G}+(d-4)(d-3)(d-2)(d-1)A^4-8(d-3)\left(R^{\mu\nu}+\frac14(d-4)Rg^{\mu\nu}\right)A_\mu A_\nu\right.\nonumber\\
&&+8(d-3)G^{\mu\nu}\nabla_\mu A_\nu+4(d-3)(d-2)\Big((d-3)A^2\divA+A^\mu\nabla_\mu A^2\Big)+4(d-3)(d-2)(\divA)^2\nonumber\\
&&-\alpha F_{\mu\nu} F^{\mu\nu}-4(d-3)(d-2)\nabla_\mu A_\nu \nabla^\nu A^\mu\Big]\,,
\label{eq:EGB}
\end{eqnarray}
with $\mathcal{G}=R_{\alpha\beta\gamma\delta}R^{\alpha\beta\gamma\delta}-4R_{\alpha\beta}R^{\alpha\beta}+R^2$ the Gauss-Bonnet term corresponding to the Levi-Civita connection and
\be
\alpha\equiv4\Big(2+4b_5-4b_6-c_2\Big)-\Big(7+4b_3-4b_4+8b_5-8b_6-3c_2-2c_3\Big)d+2\Big(1-b_2-b_4-b_6\Big)d^2
\ee
is the aforementioned only parameter (in addition to the coupling constant $\lambda$) determining the Weyl vector sector action. Since this parameter only appears as the coefficient of the usual Maxwell term of the Weyl field, it will actually appear as a coupling constant for canonically normalized fields. Moreover, in order for the Weyl field not to be a ghost, we need to impose $\lambda\alpha>0$. This is of course a necessary but not a sufficient condition for the absence of ghosts in the theory.

In addition to the previously discussed quadratic terms, we will also add the usual Einstein-Hilbert term so that our resulting action reads\footnote{Of course, we are free to add a cosmological constant, but such a term is not relevant for our discussion here.}
\be
S=\int\diff^dx\sqrt{-g}\left[-\frac12M_p^{d-2}\mR+\lambda\hG\right].
\ee
where $M_p$ is the corresponding Planck mass.  All the curvature objects present in this action are to be expressed in terms of the Levi-Civita connection and the vector field so that  we will end up with a vector-tensor theory in a (pseudo) Riemannian spacetime with the usual metric-compatible connection, that is more familiar to us.

At this point, it is worth stressing that having more geometrical objects at our disposal, one could consider more general theories. In particular, one could also introduce the very same objects, but for the Levi-Civita connection, i.e., the Einstein-Hilbert plus Gauss-Bonnet written in terms of the connection $\Gamma$. Obviously, this will not introduce any new term in the action, but it can change the corresponding coefficients. Furthermore, the Weyl part of the connection has tensorial transformation properties so that it could also be used to construct new terms. In other words, we could construct additional terms in our action by using contractions of  $R_{\mu\nu\rho}{}^\alpha$, $\hR_{\mu\nu\rho}{}^\alpha$, $\mR_{\mu\nu\rho}{}^\alpha$ and $W^\alpha_{\beta\gamma}$. Some of such contractions will be proportional to terms already present in our action, so they will only change some coefficients. In addition, we want to write our action by using purely geometric scalars  constructed out of the entire connection curvature $\mR_{\mu\nu\rho}{}^\alpha$. Finally, we have also restricted our action to non-parity violating terms, since pseudo-scalars could also be added as, e.g., contractions of the homothetic tensor and its dual of the form $\mQ^*_{\mu\nu}\mQ^{\mu\nu}$. In any case, the exhaustive exploration of all the possible terms is out of the scope of the present work, where we intend to explore the simple case of the discussed extended version of Gauss-Bonnet gravity, which already features interesting properties as we will show below.

The first term in our action and linear in the scalar curvature only contributes a mass term for the vector field when expanded in terms of non-hatted quantities,  since the last term in (\ref{expR}) is a pure divergence. Thus, if we disregard the extended Gauss-Bonnet term, we obtain the usual Einstein-Hilbert action plus a non-dynamical vector field, which, in turn, does not play any role\footnote{In that case, the field equations for the vector field are precisely $A_\mu=0$.}. Therefore, the Einstein-Hilbert action leads to the same physical gravitational theory for both purely metric and Weyl geometries and irrespective the spacetime dimension $d$. Of course, this a reflection of the well-known result that metric and Palatini formalisms lead to the same theory for the Einstein-Hilbert action, being the metric compatibility condition of the connection a consequence of the equations of motion\footnote{Actually, the Palatini formalism allows for the presence of torsion so the equivalence is only true when vanishing of torsion is imposed as an additional condition \cite{Olmo:2013lta}. It is also interesting that the equivalence of metric and Palatini formulations extends from GR to Lovelock gravities \cite{lovelock}.}.

For the extended Gauss-Bonnet term, more interesting things happen. The linear derivative coupling of $A_\mu$ in (\ref{eq:EGB}) to curvature is to the Einstein tensor so that it can be written as a pure divergence and, consequently, it only contributes as a boundary term in the action. The second term of the second line in (\ref{eq:EGB}) is also a pure divergence for $d=4$ and vanishes for $d<4$ so that it is non-trivial only in dimensions higher than $4$. At the level of the action, things get even more interesting  when performing a few integrations by parts.  We can use that $\nabla_\mu A_\nu \nabla^\nu A^\mu\rightarrow -A_\nu \nabla_\mu\nabla^\nu A^\mu=-(A_\nu \nabla_\nu\nabla^\mu A^\mu+R_{\mu\nu} A^\mu A^\nu)\rightarrow((\nabla\cdot A)^2-R_{\mu\nu} A^\mu A^\nu)$ to recast the action in the following equivalent (up to boundary terms) form:
\begin{eqnarray}
S_{\rm EGB}=\lambda\int\diff^dx\sqrt{-g}\hG&=&\lambda\int\diff^dx\sqrt{-g}\Big[\mathcal{G}+(d-4)(d-3)(d-2)(d-1)A^4+4(d-4)(d-3)G^{\mu\nu}A_\mu A_\nu\nonumber\\
&&+4(d-4)(d-3)(d-2)A^2\divA-\alpha F_{\mu\nu} F^{\mu\nu}    \Big].
%\lambda\int d^4\sqrt{-g}\left(G-12F^2\right)
\end{eqnarray}
We see how the worrisome term $(\divA)^2$ completely disappears and the non-derivative coupling to gravity gives an effective mass term determined by the Einstein tensor. Therefore, the final expression for our desired action written in terms of the Riemannian connection is
\begin{eqnarray}
S&=&\int\diff^dx\sqrt{-g}\Bigg[-\frac12M_p^{d-2}R+\lambda\mathcal{G}
+4\lambda(d-4)(d-3)(d-2)A^2\divA-\alpha\lambda F_{\mu\nu} F^{\mu\nu}  
\nonumber\\
&&
+\frac12(d-2)(d-1)M_p^{d-2}A^2+\lambda(d-4)(d-3)(d-2)(d-1)A^4+4\lambda(d-4)(d-3)G^{\mu\nu}A_\mu A_\nu
\Bigg].
\end{eqnarray}
Now, if we remind that $\alpha$ is a free parameter, we see that two branches exist in this theory depending on whether such a parameter vanishes or not. If we tune the original parameters to set $\alpha=0$, the kinetic Maxwell term $F_{\mu\nu} F^{\mu\nu} $ for the vector field disappears and the only remaining derivative term for $A_\mu$ is $A^2\divA$. Since this term only contains one derivative, it does not yield a propagating field equation for $A_\mu$ and, therefore, the Weyl vector is not a propagating degree of freedom. 

On the other hand, if we let $\alpha$ be non-vanishing, the vector field acquires a kinetic term and represents a propagating field. Although the kinetic Maxwell term has a $U(1)$ gauge invariance, the full vector sector does not, so unstable ghostly degrees of freedom might propagate. This issue will be discussed in more detail below, where we show that only 3 physical degrees of freedom actually propagate.

%This case corresponds to the choice of parameters so that $\hG$ becomes the actual Lovelock invariant, which for an arbitrary connection is given by the combination
%\be
%\mR_{\alpha\beta\gamma\delta}\mR^{\alpha\beta\gamma\delta}-\mR_{\alpha\beta}\mR^{\alpha\beta}+2\mR_{\alpha\beta}\mP^{\alpha\beta}-\mP_{\alpha\beta}\mP^{\alpha\beta}+R^2
%\ee
%i.e., it corresponds to 

In order to have a canonically normalized vector field kinetic term, we need to perform the rescaling\footnote{Notice that $\alpha\lambda$ must be positive for the transverse modes of the vector field not to be ghosts, so the rescaling is always well-defined.} $A_\mu\rightarrow \frac{1}{2\sqrt{\lambda\alpha}}A_\mu$ so that we finally get
\begin{eqnarray}
S=\int\diff^dx\sqrt{-g}\Big[-\frac12 M_p^{d-2}R(g)+\lambda\mathcal{G}-\frac14F_{\mu\nu} F^{\mu\nu} +\xi A^2\divA+\frac12 M^2A^2+\kappa A^4 +\beta G^{\mu\nu}A_\mu A_\nu \Big]\,,
\end{eqnarray}
where we have defined
\begin{eqnarray}
M^2&\equiv&\frac{(d-2)(d-1)}{4}\frac{ M_p^{d-2}}{\alpha\lambda},\\
\xi&\equiv&\frac{(d-4)(d-3)(d-2)}{2|\alpha|}(\alpha\lambda)^{-1/2},\\
\kappa&\equiv&\frac{(d-4)(d-3)(d-2)(d-1)}{16\alpha}(\alpha\lambda)^{-1},\\
\beta&\equiv&\frac{(d-4)(d-3)}{\alpha}.   
\end{eqnarray}
Notice that all the above parameters except the effective mass vanish in dimensions lower than 5,  so that 4 dimensions represents a special case that will be treated separately in the next section. There is also an interesting case when $\alpha\rightarrow\infty$ while $\alpha\lambda$ remains finite. In such a case $\beta,\kappa,\xi\rightarrow0$ and we recover the usual Proca action for a massive vector field.

Note that we restrict our considerations to classical level in this paper. By taking into account quantum loops it may be possible to obtain finite contributions from the nonmetric terms that would classically vanish at $d=4$ since, as well known, dimensional regularisation schemes generate quantum corrections to gravity scaling as $\sim 1/(d-4)$. Though there is some evidence \cite{quantum} that in the purely Riemannian context the quantum contributions from the topological Gauss-Bonnet term cancel at exactly $d=4$, to our knowledge this issue has not been investigated in the nonmetric Weyl geometry. As that lies beyond the scope of the present study, we simply take here the ''bare'' values of $\xi$, $\kappa$ and $\beta$ at their classical face value, and thus assume they vanish in $d=4$ in the next section \ref{fourd} and use dimensional reduction from $d>4$ in the following section \label{morethanfourd} to get nontrivial contributions from the three new terms.

\section{Case with $d=4$}
\label{fourd}
\label{S:d=4}
As one can see from the resulting form of the action, the 4-dimensional case is remarkable because all the non-minimal couplings of the Weyl field disappear. For this reason, we will study this simpler case separately and will leave the general case in higher dimensions for next section. We will first show how the action reduces to the Einstein-Proca theory in 4 dimensions and, then, we will discuss its potential cosmological impact, including the possibility of having dark matter from the Weyl field. We also show that a cubic Weyl term reproduces the recently studied Horndeski vector action.

\subsection{Geometrising Einstein-Proca theory}
In the four dimensional case, only the $U(1)$ gauge-invariant kinetic term for the vector field survives in the action (up to boundary terms). Thus, the Einstein Hilbert action plus the Extended Gauss-Bonnet term in a Weyl geometry in $4$ dimensions give rise to pure GR supplemented with a Proca massive vector field, i.e.:
\be \label{proca}
S=\int \diff^4x\sqrt{-g}\left(-\frac12 M_p^2\hR+\lambda\hG\right)=\int \diff^4x \sqrt{-g}\left(-\frac12 M_p^2R-\frac14 F^2+\frac12M^2A^2\right),
\ee 
where the mass of the canonically normalized vector field is given by\footnote{Notice that the EGB coupling $\lambda$ is dimensionless in $d=4$.} $M^2=\frac{3M_p^2}{2\alpha\lambda}$ and we have dropped the Gauss-Bonnet term $\lambda\mathcal{G}$ because it is topological in 4 dimensions. This corresponds to the branch of the theory with non-vanishing $\alpha$. For the branch with $\alpha=0$, the only remaining term for $A_\mu$ is the mass term coming from the Einstein-Hilbert part of the action. As explained before, in this case the vector field vanishes identically by virtue of its own field equations. It is interesting to notice that the actual Gauss-Bonnet combination (i.e., the corresponding Lovelock invariant) leads to $\alpha=0$ as it should, since it is a pure topological invariant in 4 dimensions. We note that previously Buchdahl has found a geometrisation of the Proca field by considering generalised quadratic curvature terms in the Palatini formulation \cite{Buchdahl:1979ut}.

Matter fields of the standard model of elementary particles are assumed to couple minimally to the spacetime metric. This implies that they only feel the Levi-Civita part of the spacetime connection and, thus, are insensitive to the Weyl component. For instance, the trajectory followed by a point-like particle is determined by the geodesic equations. However, the connection entering these equations as deduced from the corresponding action $S_{pp}\propto\int\diff s$ with $\diff s$ its worldline path, is actually the Levi-Civita connection of the metric. It would of course be possible to introduce non-trivial couplings of $A_\mu$ to ordinary fields by allowing non-minimal couplings, i.e., direct couplings to the connection through the curvature. This is in fact a usual result of modifying the gravitational sector. To give an example, in the model of Higgs inflation introduced in \cite{Barvinsky:2008ia}, there is a coupling of the standard model Higgs field $\Phi$ to the Ricci scalar of the form $U(\Phi)R$, with $U(\Phi)$ a certain function. In that case, an interaction term of the form $U(\Phi)A^2$ will arise from the Einstein-Hilbert term. Had we added a coupling to the EGB term, a kinetic interaction would also arise. Finally, notice that the vector field will couple via quantum loops to all the particles, but those processes are typically suppressed by the Planck scale.

Given that the Weyl vector field $A_\mu$ is decoupled from ordinary matter, it represents a natural candidate for Dark Matter (DM). It is worthwhile to notice that precisely the fact that the Weyl field is completely decoupled from the SM particles, makes it different from those based on hidden photons as in \cite{hiddenphoton} where there is a kinetic mixing between the additional massive vector boson and the usual photon. Thus, the mass constraints obtained for those models do not apply to our case. In addition, only cosmological or astrophysical constraints can be obtained for the Weyl field mass, since any potential signature in accelerators necessarily is through processes involving graviton loops, which are very suppressed as explained above. We will show below that the Weyl field is a very suitable candidate for cosmological dark matter, but first let us comment on the possibility of constructing non-trivial cubic interaction terms.

\subsection{On cubic curvature terms}

The main focus of our study so far has been to find an extension of the the Gauss-Bonnet term leading to second order equations of motion. Our approach could of course be implemented to include higher Lovelock invariants to obtain more general actions for a vector-tensor theory and with second order equations. In $d=4$, the Gauss-Bonnet term is a topological invariant and the higher Lovelock invariants vanish identically. However, in a Weyl geometry, it is possible to construct a particular combination which is cubic in the curvature and gives rise to second order field equations. Such term is allowed due to the existence of the vector-tensor Horndeski interaction given by $\Lag_{\rm H}=\frac{1}{\Lambda^2}L^{\alpha\beta\gamma\delta}F_{\alpha\beta} F_{\gamma\delta}$, with $L^{\alpha\beta\gamma\delta}\equiv-\frac12\epsilon^{\alpha\beta\mu\nu}\epsilon^{\gamma\delta\rho\sigma}R_{\mu\nu\rho\sigma}$ the double dual Riemann tensor\footnote{This type of tensor structure also appears for scalar-tensor interactions within the context of massive gravity \cite{ProxyTheory}.} and $\Lambda$ some mass scale. This term was shown to lead to second order equations of motion in \cite{Horndeski:1976gi}.  More recently, the cosmology of this type of interaction has been studied in  \cite{Barrow:2012ay}  and its stability in different non-trivial backgrounds has been analysed in \cite{Jimenez:2013qsa}.  The reason why such a combination does not lead to higher than second order equations is because the double dual Riemann tensor and the dual of $F_{\mu\nu}$ are both identically divergenceless. Guided by the existence of this interaction, we can construct particular cubic terms in $\mR_{\mu\nu\rho}{}^\sigma$ that reduce to the Horndeski vector-tensor interaction and give a trivial contribution for the pure Levi-Civita connection. The desired combination is given by 
\be
\Lag_{\rm H}=-\frac{1}{4 \Lambda^2}\hat{L}^{\alpha\beta\gamma\delta}\mQ_{\alpha\beta} \mQ_{\gamma\delta}\,,
\ee
where $\hat{L}^{\alpha\beta\gamma\delta}\equiv-\frac12\epsilon^{\alpha\beta\mu\nu}\epsilon^{\gamma\delta\rho\sigma}\mR_{\mu\nu\rho\sigma}$ denotes the double dual Riemann tensor of the full connection and $\mQ_{\mu\nu}$ the homothetic tensor. This gives our desired contribution because, for the Weyl connection, we have $\mQ_{\mu\nu}=-4F_{\mu\nu}$. Notice that we could have added combinations including $\mP_{\mu\nu}$, although in all cases only its antisymmetric part will contribute and this part is precisely fully determined by $F_{\mu\nu}$, as we explained in detail above. It is also worth stressing that the double dual Riemann tensor is still divergenceless in a Weyl geometry (which is the key property that guarantees the second order nature of the field equations), but it is divergenceless with respect to the full connection $\hat{\Gamma}^{\alpha}_{\beta\gamma}$ and not with respect to the Levi-Civita part of the connection, i.e., $\hnabla\cdot L=0$, but $\nabla\cdot L\neq0$.

The discussed type of Horndeski vector-tensor interactions arise from higher dimensional scenarios when performing a Kaluza-Klein reduction of the higher dimensional Gauss-Bonnet term as interaction between the graviphoton and the 4-dimensional graviton \cite{KKreduction}. In Weyl geometry, we see that these interactions can be obtained from geometrical objects directly in 4 dimensions as cubic curvature terms.

Written in terms of the Levi-Civita connection, Riemannian objects and the vector field, the cubic term leads to the following vector-tensor interactions
\be
\Lag_H=\frac{1}{\Lambda^2}\Big[L^{\alpha\beta\gamma\delta}F_{\alpha\beta} F_{\gamma\delta}
+4\Big(2F_{\alpha}{}^\mu F^{\alpha\nu}-F_{\alpha\beta}F^{\alpha\beta}g^{\mu\nu}\Big)\nabla_\mu A_\nu-8F_{\alpha}{}^\mu F^{\alpha\nu}A_\mu A_\nu
\Big].
\label{LagH}
\ee
Here we explicitly see how the proposed cubic term leads to the Horndeski type of vector-tensor interaction plus some additional terms all of which yield second order equations for both, the Weyl field and the metric tensor. As we have commented, the stability of the pure Horndeski term was studied in detail in \cite{Jimenez:2013qsa}. In that work, only gauge invariant terms were considered so that the vector field only propagates 2 degrees of freedom. However, $U(1)$ gauge invariance is absent in (\ref{LagH}) so additional modes might propagate. It is not our aim to study the full stability of the modified Horndeski lagrangian (\ref{LagH}) here, but let us simply point out the following. If we consider the pure longitudinal mode $A_\mu=\partial_\mu\phi$, the action trivializes because $F_{\mu\nu}$ vanishes. If transverse modes contributing to $F_{\mu\nu}$ are present, then the lagrangian for the pure longitudinal mode becomes 
\be
\Lag_{\phi}=\frac{1}{\Lambda^2}\Big(K^{\mu\nu}\nabla_\mu\nabla_\nu\phi+M^{\mu\nu}\partial_\mu\phi\partial_\nu\phi\Big)\,,
\ee
where we have defined the symmetric objects
\begin{eqnarray}
K^{\mu\nu}&\equiv& 4\Big(2F_{\alpha}{}^\mu F^{\alpha\nu}-F_{\alpha\beta}F^{\alpha\beta}g^{\mu\nu}\Big)\,, \\
M^{\mu\nu}&\equiv& -8F_{\alpha}{}^\mu F^{\alpha\nu}\,,
\end{eqnarray}
which only depend on $F_{\mu\nu}$. This lagrangian for $\phi$ leads to contributions of second order derivatives in $\phi$, which is an indication that no more than 3 propagating degrees of freedom are present in the theory, as it corresponds to a Proca field in 4 dimensions. Of course, this does not guarantee the full stability of the theory since instabilities and/or lack of hyperbolicity in the field equations can exist in vector field theories propagating fewer than 4 degrees of freedom \cite{vectorinstabilities}. To end this discussion on cubic curvature terms, let us notice that cubic terms involving only the homothetic tensor of the form $\mQ^\alpha{}_\beta \mQ^\beta{}_\gamma\mQ^\gamma{}_\alpha$ vanish identically so they do not give new contributions.

\subsection{Cosmological impact: A natural Dark Matter candidate}
The mass of the vector field is fully determined by the combination $\alpha \lambda$ so that its cosmological relevance and consistence as DM model will constrain such a parameter. If we assume a {\it natural} value for it, i.e., that it takes a value $\alpha\lambda\sim 1$, then the vector field has a mass close to the Planck scale and it represents an extremely heavy field whose interaction range is below Planck length. Thus, throughout the cosmological evolution of the universe, the homogeneous part of the spacelike components of the vector field will be in a rapidly oscillatory regime, giving rise to an averaged energy density decaying like a matter fluid. To see this, let us write the vector field equations in a FLRW universe, whose metric is given by
\be
\diff s^2=\diff t^2-a(t)^2\diff\vec{x}^2.
\ee
The equation for the temporal component gives the usual constraint $A_0=0$. For the spatial part of the homogeneous vector field, the three components evolve independently of each other and satisfy the equation
\be
\ddot{A}_i+H\dot{A}_i+M^2A_i=0.
\ee
From this equation we can clearly see that the field components will oscillate with frequency $\omega\sim M$ throughout most of the expansion history of the universe under the assumption $\alpha\lambda\sim 1$. In general, the field will enter the oscillatory regime whenever $H^2 <M^2=3M_p^2/(2\alpha\lambda)$. These oscillations can be averaged and satisfy the virial relation 
\be
\left\langle\frac{\dot A_i \dot A_j}{a^2} \right\rangle=M^2\left\langle\frac{A_iA_j}{a^2}\right\rangle,
\ee
where the brackets denote average over several oscillations. The trace of this relation yields $\big\langle\dot {\vec{A}}^2/a^2\big\rangle=\big\langle M^2A_i^2/a^2\big\rangle$. The energy-momentum components of the vector field in a FLRW metric are
\begin{eqnarray}
 \rho &\equiv& T^0_{\;\;0}= \frac{1}{2a^2} \Big(\dot {\vec{A}}^2  +M^2\vec{A}^2\Big)\,, \\
p_k &\equiv& - T^k_{\;\;k } = \frac{1}{2a^2}\Big(\dot {\vec{A}}^2  - 2\dot{A}_k \dot{A}_k - M^2\vec{A}^2 +2M^2 A_k A_k\Big), \; k=1,2,3\,,
\\
T^i_{\;\;j} &=& \frac{1}{a^2}\Big(\dot{A}_i \dot{A}_j -M^2A_i A_j\Big), \;\;\; i\neq j \,.
\end{eqnarray}
Thus, we can see that the averaged pressures and anisotropic stresses $\langle p_k\rangle$ and $\langle T^i{}_j\rangle$ vanish and, as a consequence, the averaged energy momentum tensor of the vector field will evolve as that of a matter component. Notice that this is independent of the polarization state (linear, circular...) of the vector field. Hence, the coherent oscillations of this vector field can play the role of a dark matter component.  In fact, this result is nothing but a particular case of the more general case considered in \cite{Cembranos:2012kk} with an arbitrary power law potential and subsequently generalized for non-abelian fields in \cite{Cembranos:2012ng} and for arbitrary spin fields in \cite{Cembranos:2013cba}. 

We can obtain constraints on the initial amplitude of the vector field for it to be compatible with the observed dark matter abundance today which is $\Omega_{DM}\simeq0.2$ so we need to impose\footnote{Notice that $\rho_0$ denotes the energy density of the vector field today, not the critical density.} 
\be
\left\langle\frac{8\pi G\rho_0}{3H_0^2}\right\rangle\lsim\Omega_{DM}.
\ee
Since we have that the average energy density evolves as that of a matter component, we can relate the energy density today to the initial amplitude of the vector field as $\langle\rho_{ini}\rangle=\langle\rho_0\rangle(1+z_{ini})^{3}=-M^2\langle A^2_{ini}\rangle(1+z_{ini})^{3}$ with $A^2_{ini}=-\vec{A}^2/a^2\vert_{ini}$ and we have used the averaged virial relation. Thus, the above constraint can finally be written as
\be
\langle-A_{ini}^2\rangle\lsim 2\alpha\lambda\Omega_{DM}(1+z_{ini})^3H_0^2\,,
\ee
where we have expressed the mass of the vector field in terms of the original parameters $\lambda$ and $\alpha$. If these parameters are of order 1, then we roughly have the bound for the initial amplitude of the vector field $\sqrt{\langle-A_{ini}^2\rangle}\lsim10^{-33}(1+z_{ini})^{\frac32}$eV, which is very small value for most of the universe history. For instance, at BBN we have $\sqrt{\langle-A_{BBN}^2\rangle}\lsim10^{-18}$eV. Thus, we would need a mechanism to generate the field giving rise to a small primordial amplitude. This can require a fine tuning of the initial conditions. One possibility would be from quantum fluctuations during inflation, which would deserve a more detailed analysis.

\section{Case with $d>4$}
\label{morethanfourd}
In dimensions higher than $4$, the action becomes much more involved and, in particular, the vector field acquires non-minimal couplings. In this section, we will study the stability of the theory and will show how one can recover Horndeski type of interactions.

\subsection{Degrees of freedom}
A worrisome feature of the resulting action for the Weyl vector field is the absence of a $U(1)$ gauge invariance in derivative terms, which might potentially lead to the propagation of additional (ghostly) degrees of freedom with respect to the 3 healthy degrees of freedom corresponding to a massive vector field. However, this is not the case for the Weyl field, as can be seen from several ways, but all essentially reduces to the fact that no proper kinetic term is present in the action for $A_0$. First of all, one should notice that, although all four components of the vector field enter with time derivatives in the action, the corresponding Hessian vanishes because
\be
\frac{\delta^2 \Lag}{\delta \dot{A}_0\delta \dot{A}_\mu}=0,
\ee
which indicates a degenerate transformation to canonical hamiltonian variables and the presence of constraints. Such constraints can be obtained from the definition of the conjugate momenta, which are given by 
\begin{eqnarray}
&&\Pi^0=\xi A^2,\\
&&\Pi^i=F^{i0}.
\end{eqnarray}
From this, we see that there is one primary constraint obtained from the canonical momentum of the temporal component given by $\Pi^0-\xi A^2=0$. It is easy to see that such a primary constraint generates a secondary constraint via its Poisson bracket with the Hamiltonian. Moreover, this secondary constraint has a non-vanishing Poisson bracket with the primary constraint, which shows that it corresponds to a second class constraint and, thus, it does not generate any gauge symmetry, as expected from the obvious absence of $U(1)$ gauge invariance in the action. This hints that only the 3 physical degrees of freedom corresponding to a massive vector field propagate. Obviously, showing that no additional degrees of freedom are present does not guarantee the stability of the theory, since we could still have a ghostly longitudinal mode or other types of instabilities, as we discussed in the 4-dimensional case \cite{vectorinstabilities}. However, having only 3 propagating modes  prevents the appearance of the Ostrogradski ghost associated to the higher order nature of the St\"uckelberg field, as we discuss in more detail in the following.

\subsection{Recovering Horndeski Lagrangians}
We will use the St\"uckelberg trick to have a $U(1)$ gauge symmetry for the vector field. This is achieved, as usual, by making the replacement\footnote{In this subsection we will set $d=5$ for simplicity. Then, we will use capital latin indices for 5 dimensional objects whereas greek indices will be reserved for 4 dimensional objects.} $A_N\rightarrow A_N+\frac 1M\partial_N\phi$ in the action. This is of course equivalent to making the Weyl connection $U(1)$ invariant by performing said replacement in the non-metricity condition (\ref{nonmetricity}). This is useful because one can see that the absence of the $U(1)$ symmetry  for the vector field leads to the appearance of a ghost degree of freedom as a consequence of having higher than second order derivative equations for the St\"uckelberg field with the associated Ostrogradski instability. When the St\"uckelberg field only propagates one degree of freedom (i.e., it satisfies second order equations), it represents the longitudinal mode of the vector field. After the mentioned replacement, in our case, the resulting $U(1)$ invariant lagrangian can be split into the sum of 3 pieces $\Lag=\Lag_\phi+\Lag_A+\Lag_{int}$ corresponding to the pure scalar and Weyl fields and the term mixing both of them. The scalar field piece can be written as:
%\begin{eqnarray}
%\Lag_{\phi}=\frac12(\partial\phi)^2\left[1+2\lambda\frac{(d-4)(d-3)}{(d-2)(d-1)}\frac{(\partial\phi)^2}{M_p^2}\right]+4\lambda\frac{(d-4)(d-3)}{\sqrt{(d-2)(d-1)^3}}\frac{(\partial\phi)^2\Box\phi}{M_p^3}+4\lambda\frac{(d-4)(d-3)}{(d-2)(d-1)}\frac{G^{\mu\nu}\partial_\mu\phi\partial_\nu\phi}{M_p^2}
%\end{eqnarray}
%\begin{eqnarray}
%\Lag_\phi=\frac12(\partial\phi)^2\left(1+\frac{2\alpha}{M^4}(\partial\phi)^2\right)+\frac{\kappa}{M^3}(\partial\phi)^2\Box\phi+\frac{\beta}{M^2}G^{AB}\partial_A\phi\partial_B\phi
%\end{eqnarray}
\begin{eqnarray}
\Lag_\phi=\frac12\left[\left(1+\frac{2\kappa}{M^4}(\partial\phi)^2\right)g^{AB}+\frac{2\beta}{M^2}G^{AB}\right]\partial_A\phi\partial_B\phi+\frac{\xi}{M^3}(\partial\phi)^2\Box\phi.
\end{eqnarray}
where we can recognize the Horndeski form, thus, leading to second order equations of motion and avoiding the Ostrogradski ghost. In the case of a Weyl geometry fully determined by the scalar field (i.e., with $A_\mu\equiv\partial_\mu\phi$), we recover exactly this particular type of Horndeski action, without the vector sector\footnote{Notice that if the Weyl field is assumed to be purely longitudinal, i.e., $A_\mu\equiv\partial_\mu\phi$, the non-metricity condition reduces to a pure conformal transformation}. Of course, this Horndeski action is only non-trivial in dimensions higher than 4. As we discussed above, in $d=4$ only the canonical kinetic term for the scalar field survives. 

%\subsection{KK reduction}
In order to obtain non-trivial actions in 4 dimensions, we need to use some method to arrive at the effective 4-dimensional action. We want to explicitly show how to obtain 4-dimensional Horndeski terms from Extended Gauss-Bonnet gravity in Weyl geometry so that we will use the simplest method and perform a standard Kaluza-Klein (KK) reduction assuming that we have the usual 4 dimensions plus one compact fifth dimension. In general, the reduction procedure will lead to the appearance of an infinity tower of KK Horndeski fields which are non-trivially coupled to each other. Nevertheless, the scalar field originates from a St\"uckelberg trick so that it has a shift symmetry\footnote{In the full action including the Weyl vector field, $\phi$ has a local shift symmetry introduced to guarantee the $U(1)$ symmetry for the vector field. This allows to gauge away the St\"uckelberg field. However, in this section we are focusing on the scalar field itself and we ignore the vector field because our aim is to recover Horndeski terms.}, i.e., all the interactions are derivative (as one can straightforwardly check by looking at its lagrangian). This protects the zero mode of the tower from non-derivative interactions with the higher modes. In fact, it is the zero mode that inherits the shift symmetry of the 5-dimensional field. If we send the extra dimension radius to a very small scale, the KK reduction procedure yields the following effective 4 dimensional lagrangian
\be
\Lag_\phi=\sum_n\left[\frac12\partial_\mu\varphi_n\partial^\mu\varphi_{-n}+\frac{\beta}{M^2}G^{\mu\nu}\partial_\mu\varphi_n\partial_\nu\varphi_{-n}+\frac{\xi}{M^3}\sum_m\partial_\mu\varphi_n\partial^\mu\varphi_m\Box\varphi_{-(n+m)}+\frac{\kappa}{M^4}\sum_{m,p}\partial_\mu\varphi_n\partial^\mu\varphi_m\partial_\mu\varphi_p\partial^\mu\varphi_{-(m+n+p)}\right]
\ee
with $\varphi_n$ the corresponding KK modes of $\phi$. Now, we can see that the pure zero mode exactly reproduces a Horndeski type of lagrangian. This is the simplest reduction to obtain the Horndeski lagrangians where we have set the dilaton (and the graviphoton) to zero. For the mentioned zero mode, (i.e. if we assume that $\phi$ only depends on the 4-dimensional coordinates), the action for the St\"uckelberg field resembles an extended version of Kinetic Gravity Braiding \cite{KGB} whose general lagrangian is\footnote{This is of course a particular case of the general Horndeski lagrangian. See Section \ref{s:weylhorndeski}.} $\Lag_{\rm KGB}=K(\phi,X)+G(\phi,X)\Box\phi$ with $X\equiv\frac12\partial_\mu\phi\partial^\mu\phi$.  After integrating by parts, the action for the zero mode can be written as
\begin{eqnarray}
\Lag_{\varphi_0}=\frac12\left[1+\frac{2\kappa}{M^4}(\partial\varphi_0)^2\right](\partial\varphi_0)^2-\left[\frac{\beta}{M^2}\varphi_0G^{\mu\nu}-\frac{\xi}{M^3} (\partial\varphi_0)^2g^{\mu\nu} \right]\nabla_\mu\nabla_\nu\varphi_0
\end{eqnarray}
where we explicitly see the aforementioned extended form of the Kinetic Gravity Braiding model. 

If one includes also the vector part of the nonmetricity, the following interaction lagrangian emerges after the KK reduction to four dimensions:
\bea
\Lag_{int}& = &MA\cdot\nabla\varphi_0 + \frac{2\kappa}{M}\lb 2A^2 A\cdot\nabla\varphi_0 + \frac{1}{M}(\partial\varphi_0)^2
\lp A^2 + \frac{1}{M} A\cdot\nabla\varphi_0\rp + \frac{2}{M}(A\cdot\nabla\varphi_0)^2\rb 
\nonumber \\
& + & \frac{\beta}{M^2}G^{\mu\nu}A_\mu\nabla_\nu\varphi_0 + \frac{\xi}{M^3}(\partial\varphi_0)^2\Box\varphi_0\,.
\eea
The suitability of these vector-scalar theories to cosmological applications along the lines of \cite{vectorDE,vectorInflation,vectorcurvaton,vectorMG} as well as their full stability remains to be studied. In the following we will instead consider conformally invariant versions of the Horndeski-type theories we have arrived at here.

\section{Weyl gauge scale invariant theory}
\label{s:weylhorndeski}
In the previous section, we have shown how particular types of Horndeski interactions arise for the longitudinal mode (or St\"uckelberg field) of the Weyl vector field within our extended Gauss -Bonnet action. So far, we have not considered the fact that Weyl geometries are the natural arena to construct theories with a local scale invariance or conformal gauge symmetry\footnote{While this work was being completed, we came accross Ref. \cite{Padilla:2013jza}, where Weyl gauging was also used to make the Horndeski lagrangians scale-invariant. In that work, the authors also generalize the results to multiscalar theories. Our results coincide with theirs where they overlap.}. In fact, the vector field is nothing but the gauge field associated to said symmetry as we showed in Section 2. 

Let us briefly review how the Weyl gauging works for a scale invariant theory. For that, we need a conformal symmetry for the metric $g_{\mu\nu}\rightarrow e^{2\Lambda}g_{\mu\nu}$. Then, one can build the action by means of the Weyl tensor, which is conformally invariant, but it introduces ghostly degrees of freedom. We follow another route here. One can gauge the conformal symmetry by introducing a vector field to construct a covariant derivative as
\be
\mD_\mu g_{\alpha\beta}=(\partial_\mu+2A_\mu)g_{\alpha\beta}.
\ee
Of course, this covariant derivative transforms in the correct way $\mD_\mu g_{\alpha\beta}\rightarrow e^{2\Lambda}\mD_\mu g_{\alpha\beta}$ provided the gauge field transforms as $A_\mu\rightarrow A_\mu-\partial_\mu\Lambda(x)$, i.e., as a $U(1)$ gauge field. Now, to  determine the connection, one can impose the generalised metric condition
\be
\D_\mu g_{\alpha\beta}\equiv (\hnabla_\mu+2A_\mu)g_{\alpha\beta}=0\,.
\ee
It is not very difficult to understand that the connection is then given by
\be
\hGam^\alpha_{\beta\gamma}=\frac12 g^{\alpha\lambda}\left(\mD_\beta g_{\lambda\gamma}+\mD_\gamma g_{\beta\lambda}-\mD_\lambda g_{\beta\gamma}\right)\,,
\label{connection}
\ee
since $\D=\mD+\hGam$. This is nothing but expression (\ref{Wconnection}) written in a manifestly Weyl invariant form. From the way the connection is written in (\ref{connection}), it is  obvious that it is a conformal gauge invariant object. Therefore, the corresponding Riemann tensor $\mR_{\mu\nu\rho}{}^\sigma$ constructed out of this connection will also be invariant and,  thus, the Ricci scalar will transform as the inverse metric, i.e., $\mR\rightarrow e^{-2\Lambda}\mR$. That way, the usual Einstein-Hilbert action for this connection can be made scale invariant if we introduce a scalar field playing the role of the Planck mass as follows
\be
S_{EH}=-\frac\alpha2\int\diff^4x\sqrt{-g}\phi^2\mR\,,
\ee
where the scalar field transforms  as $\phi\rightarrow e^{-\Lambda}\phi$. Of course, now one could also include a gauge invariant kinetic term for the scalar field of the form $\sqrt{-g}g^{\mu\nu}\mD_\mu\phi \mD_\nu \phi$ with $\mD_\mu \phi\equiv(\partial_\mu-A_\mu)\phi$ and the scale invariant potential $\nu\sqrt{-g}\phi^4$. On the other hand, we can also add all the terms satisfying the usual conditions plus scale invariance, that is, we can write down all the scale invariant terms leading to second order equations of motion. Such terms are
\be
S=\int\diff^4x\sqrt{-g}\left[-\frac\alpha2 \phi^2\mR+\frac12g^{\mu\nu}\mD_\mu\phi \mD_\nu \phi-\nu\phi^4+\lambda\hG-\frac14F^2\right]\,.
%+\frac{\beta}{\phi^2}\hL^{\alpha\beta\gamma\delta}F_{\alpha\beta}F_{\gamma\delta}\,.
\ee
In terms of the metric connection, this action can be written as:
\be
S=\int\diff^4x\sqrt{-g}\left[-\frac\alpha2 \phi^2R+\frac12g^{\mu\nu}\mD_\mu\phi \mD_\nu \phi-\nu\phi^4-\frac14F^2+3\tilde{\alpha}\phi^2 A^2+3\alpha\phi^2\divA\right]\,.
%+\frac{\beta}{\phi^2}\hL^{\alpha\beta\gamma\delta}F_{\alpha\beta}F_{\gamma\delta}\,.
\ee
Notice that $\mD_\mu$ acting on the scalar field does not depend on the spacetime connection, as usual. These are the usual terms quadratic in  derivatives. However, these are not the only terms leading to second order equations of motion, but they are given by the general Horndeski \cite{Horndeski:1974wa}
\begin{eqnarray}
&&\Lag_2=K(\phi,X)\,,\\
&&\Lag_3=G_3(\phi,X)\Box\phi\,,\\
&&\Lag_4=G_4(\phi,X)R-G_{4,X}(\phi,X)\Big[(\Box\phi)^2-(\nabla_\mu\nabla_\nu\phi)^2\Big]\,,\\
&&\Lag_5=G_5(\phi,X)G_{\mu\nu}\nabla^\mu\nabla^\nu\phi+\frac16G_{5,X}(\phi,X)\Big[(\Box\phi)^3-3(\Box\phi)(\nabla_\mu\nabla_\nu\phi)^2+2(\nabla_\mu\nabla_\nu\phi)^3\Big]\,,
\end{eqnarray}
with $K$, $G_3$, $G_4$, $G_5$ arbitrary functions and $X=\frac12\partial_\mu\phi\partial^\mu\phi$. The special structure of these 5 terms guarantees that the equations of motion will remain of second order even though second order derivatives are present. Now, we can wonder if these terms can be made scale invariant without spoiling such a structure. This can indeed be done in a straightforward way by simply replacing covariant derivatives by the corresponding Weyl gauge covariant derivatives, i.e., we shall replace $\nabla_\mu\rightarrow\mD_\mu$.  For the following it will be useful to remind that $[\phi]_W=-1$, $[g_{\mu\nu}]_W=2$ and $[X]_W=-4$, where $[\cdot]_W$ denotes the Weyl weight of the corresponding quantity. In order not to spoil the scale invariance of the theory, the arbitrary functions appearing in the Horndeski lagrangians must be of a determined form. For instance, for $\Lag_2$ to lead to a scale invariant term, we should require $K(\phi,X)$ to have weight\footnote{Each lagrangian should have weight -4 to compensate for the weight of the determinant in the integration measure so that the action has weight 0.} -4 . Then, we can simply require that $K$ should be of the form $\phi^4 k(\hX)$ with $\hX\equiv X/\phi^4$. If this function admits an expansion around $\hX=0$, we have $\Lag_2=\phi^4(k(0)+k'(0)X/\phi^4 +\Od(\hX^2))$ so we recover the quadratic kinetic term and the quartic potential. We can proceed analogously with $\Lag_3$ and, then, we find that $G_3$ must have weight $-1$ so that it should be of the form\footnote{Notice that we are {\it correcting} the weight of the functions by adding an appropriate power of the scalar field. One could of course choose any other combination of fields, but all of them will be equivalent to our choice by a simple redefinition of the weightless function.} $G_3=\phi g_3(\hX)$. We can proceed analogously for $G_4$, although some care must be taken to guarantee the special structure of $\Lag _4$. Obviously, $G_4$ should have weight $-2$ so that we can write $G_4=\phi^2 g_4(\hX)$.  On the other hand, for the second term in $\Lag_4$ to be scale invariant, we should demand $G_{4,X}$ to have weight $2$. This is indeed the case, since $G_{4,X}=\partial_X(\phi^2g_4(\hX))=g'_4(\hX)/\phi^2$, that has weight 2 (remember that $g_4$ and $\hX$ are scale invariant by construction). We can proceed similarly with the fifth Horndeski term to find that $G_5$ must be of the form $G_5=g_5(\hX)$.
Thus, the scale-invariant version of the Horndeski terms are given by
\begin{eqnarray}
&&\Lag_2=\phi^4k(\hX)\,,\\
&&\Lag_3=g_3(\hX)\phi\D^2\phi\,,\\
&&\Lag_4=g_4(\hX)\phi^2\mR-g'_{4}(\hX)\Big[(\D^2\phi)^2-(\D_\mu\D_\nu\phi)^2\Big]\,,\\
&&\Lag_5=g_5(\hX)\hat{G}_{\mu\nu}\D^\mu\D^\nu\phi+\frac16g'_{5}(\hX)\frac{(\D^2\phi)^3-3(\D^2\phi)(\D_\mu\D_\nu\phi)^2+2(\D_\mu\D_\nu\phi)^3}{\phi^2}\,,
\end{eqnarray}
where $\D^2\equiv g^{\mu\nu}\D_\mu\D_\nu$ and a prime denotes derivative with respect to $\hX$. In 4 dimensions, the higher Lovelock invariants are not included because the Gauss-Bonnet term is topological and the higher invariants are trivial. However, being in a Weyl geometry, we have shown that higher non-trivial terms can be introduced. In particular, we can add our extended Gauss-Bonnet term, that will simply contribute a Maxwell term for the Weyl vector (see Section \ref{S:d=4}). Since quadratic terms in curvature are automatically scale invariant, we do not need to introduce a coupling to the scalar field to correct for the weight. On the other hand, we can also include the cubic term discussed in Section  \ref{S:d=4} which leads to the vector-tensor Horndeski interaction. That term has Weyl weight $-2$ so that we need to introduce a coupling to $\phi^{-2}$ to make it scale invariant. The resulting term with conformal invariance is given by
\be
\Lag_{\rm WH}=-\frac{1}{4 \phi^2}\hat{L}^{\alpha\beta\gamma\delta}\mQ_{\alpha\beta} \mQ_{\gamma\delta}\,.
\ee
Thus, the Weyl invariant Horndeski scalar-tensor theory is given by a combination of the above 5 Horndeski lagrangians plus our extended Gauss-Bonnet term (which in 4 dimensions simply gives a Maxwell term for the Weyl field) and the extended scale invariant vector-tensor Horndeski interaction.

\section{Discussion}
\label{conclusions}
In this work we have explored extensions of General Relativity within the framework of Weyl geometries. We have started by very briefly reviewing the main geometrical properties of a Weyl spacetime, which is characterized by a non-metricity tensor fully determined by a vector field. Then, we have considered theories up to quadratic terms in the curvature under the condition that the resulting action should reduce to the usual Gauss-Bonnet term when imposing metric compatibility, i.e., when we have vanishing Weyl connection. The reason to impose such a condition was to guarantee the absence of ghosts associated to higher order derivative terms for the metric tensor. We have also restricted our action to be constructed with the total Riemann curvature tensor and discussed the independent contractions that can be performed. This procedure led us to an effective vector-tensor theory of gravity in a more familiar (pseudo) Riemannian geometry. The resulting theory is remarkably simple given our starting point and only depends on 2 parameters (in addition to the Planck mass), one of which controls whether the Weyl vector propagates or not.  The terms which survive in the action crucially depend on the number of dimensions and we have analysed separately the cases $d=4$ and $d>4$. For $d=4$, we have obtained a simple Proca action for the Weyl vector and have proposed that it could represent a natural candidate for dark matter, although the primordial amplitude must be very small. Moreover, we have discussed a cubic term in the curvature that nevertheless leads to second order field equations and which is motivated by the vector-tensor Horndeski interaction.

Thus,  in the nonmetric context of Weyl geometry, we have discovered novel geometric interpretations for several familiar theories. To summarise our findings,
\begin{itemize}
\item The Proca action is an Einstein-Gauss-Bonnet theory of nonmetric curvature. We proposed that the massive vector in this case could play the role of cosmological dark matter.  
\item Geometrisation of massless Maxwell theory on the other hand is provided by considering nonmetricity solely in the Gauss-Bonnet-like term\footnote{Alternatively one can take formally the limit $\lambda\rightarrow\infty$ in the fully nonmetric action (\ref{proca}).}. By coupling then a complex scalar field to the nonmetric covariant derivative one obtains a version of scalar electrodynamics, though we didn't explore this possibility.
\item A cubic curvature term was shown to reproduce the Horndeski vector theory with a gauge-invariance breaking generalisation.
\item The dimensionally reduced $d>4$ theories feature a scalar degree of freedom with a general Horndeski action. 
\item Finally, we considered endowing the Horndeski theories with local scale invariance by restricting to those forms of the terms that respect the scale transformation symmetry. 
\end{itemize}
Since we have considered extended Gauss-Bonnet actions in Weyl geometries, it was natural to study general scalar-tensor theories with conformal invariance because the Weyl field can be associated to the gauge field of local conformal symmetry. Thus, we have taken the most general scalar-tensor theory with second order equations of motion, i.e. the Horndeski lagrangians, and used the Weyl field to provide it with local conformal invariance. Once again, the fact that we are considering Weyl geometries has allowed us to add more terms yielding second order equations of motion and which trivialize for Riemannian geometries. Such terms are our extended Gauss-Bonnet and vector-tensor Horndeski interactions. Whereas the extended Gauss-Bonnet simply gives a Maxwell term for the Weyl vector, the extended Horndeski term gives new interactions involving the scalar field.

	In this work we have shown how vector-tensor theories of gravity can be naturally accommodated within the context of Weyl geometry. Some of the interactions are novel in the literature and allow to recover particular classes of Horndeski lagrangians with a potential interesting phenomenology for cosmological applications. These will be explored in more detail in a future work to test the true observational viability of such models. Another aspect that deserves further exploration would be to find gravitational actions with a local scale invariance which reduce to our extended Gauss-Bonnet terms in the low energy limit after a symmetry breaking mechanism so that our dimensionful parameters are promoted into expectation values of some fields. As stressed several times, we have restricted the analysis to our simple extension of Gauss-Bonnet terms and a more exhaustive exploration allowing for more general terms, including parity-violating terms, would deserve attention.

\acknowledgments

We would like to thank Antonio Maroto for useful comments on the manuscript and Lavinia Heisenberg and Gonzalo J. Olmo for useful discussions.
J.B.J. is supported by the Wallonia-Brussels Federation grant ARC No.~11/15-040 and also thanks support from the Spanish MICINN Consolider-Ingenio 2010 Programme under grant MultiDark CSD2009-00064 and project number FIS2011-23000.

\end{document}